%% file: attRec.tex
\documentclass[conference]{IEEEtran}
\IEEEoverridecommandlockouts

\usepackage{amssymb, multirow}

\input{preamble}

\begin{document}

\title{A Neural Attention Model for Adaptive Learning of Social Friends' Preferences}

\author{\IEEEauthorblockN{Dimitrios Rafailidis}
\IEEEauthorblockA{\textit{Maastricht University} \\
Maastricht, The Netherlands \\
dimitrios.rafailidis@maastrichtuniversity.nl}

\and

\IEEEauthorblockN{Gerhard Weiss}
\IEEEauthorblockA{\textit{Maastricht University} \\
Maastricht, The Netherlands \\
gerhard.weiss@maastrichtuniversity.nl}
}

\maketitle

\begin{abstract}
Social-based recommendation systems exploit the selections of friends to combat the data sparsity on user preferences, and improve the recommendation accuracy of the collaborative filtering strategy. The main challenge is to capture and weigh friends' preferences, as in practice they do necessarily match. In this paper, we propose a Neural Attention mechanism for Social collaborative filtering, namely NAS. We design a neural architecture, to carefully compute the non-linearity in friends' preferences by taking into account the social latent effects of friends on user behavior. In addition, we introduce a social behavioral attention mechanism to adaptively weigh the influence of friends on user preferences and consequently generate accurate recommendations. Our experiments on publicly available datasets demonstrate the effectiveness of the proposed NAS model over other state-of-the-art methods. Furthermore, we study the effect of the proposed social behavioral attention mechanism and show that it is a key factor to our model's performance. 
\end{abstract}

\begin{IEEEkeywords}
Neural attention models, deep collaborative filtering, social relationships, adaptive learning
\end{IEEEkeywords}

\section{Introduction}\label{sec:intro}
The collaborative filtering strategy has been widely adopted in recommendation systems~\cite{Kor09}. Matrix factorization techniques are representative collaborative filtering strategies, which factorize the data matrix with user preferences, to reveal the latent associations between users and items~\cite{PMF}. Notice that matrix factorization strategies belong to the generic class of non-linear dimensionality reduction techniques~\cite{RafailidisNM11}. In the real-world scenario the data sparsity of user preferences degrades the recommendation accuracy, as users select a few items and there are only a few preferences on which to base the recommendations. So, one of the motivations of this paper is concerned with estimating how much data sparsity affects the performance of recommendation systems and with finding solutions to improve the performance of collaborative filtering in the presence of data sparsity. In the literature several methods have been proposed to overcome this problem. For instance,~\cite{KDD09,SOCIALMF,WSDM,STE} describe methods that exploit the selections of social friends in collaborative filtering, accounting for the fact that social friends have similar preferences. However, the preferences of social friends do not necessarily match, having different influence when generating recommendations~\cite{SPF}. From the user feedback we have to learn both about her preferences and how much she is affected by her friends' selections. Nonetheless, this is a challenging task as it is difficult to determine how much her preferences are influenced by her friends when generating recommendations. In the context of social regularization in recommendation systems, recent studies try to capture the complex relationships between user preferences and those of her social friends~\cite{SPF}. In an attempt to compute the non-linearity in friends' preferences, a few deep learning strategies have been introduced to generate recommendations with social relationships, such as feedforward neural networks~\cite{FanLC18}, Denoising Autoencoders~\cite{DengHXWW17} and Restricted Boltzmann Machines~\cite{NguyenL16}. 

However, what is missing from existing studies is that we have to learn how to adaptively select a subset of friends with similar behavior and weigh their influence accordingly to improve the recommendation accuracy. Recently,  attention mechanisms have been shown to be effective in various tasks such as image captioning~\cite{Xu15} and machine translation~\cite{Bah15}, among others.  Essentially the idea behind such mechanisms is that the outputs of neural models depend on \emph{`relevant'} parts of some input that the models should focus on. Armed with different attention mechanisms, various recommendation models have been designed to generate sequential~\cite{Liu18,Man18,Sun18} and context-aware recommendations~\cite{Tay18,Xiao17}. Nonetheless, these attention models are introduced to either perform next-item recommendation or focus on users' contextual factors, and do not learn the influence of social friends.

To overcome the shortcomings of existing social recommendation strategies, we propose a Neural Attention model to adaptively learn the influence of Social friends on user preferences, namely NAS, making the following contributions: (i) \emph{We design a neural architecture to carefully capture the non-linear associations between the preferences of a user and those of her friends, by taking into account the social latent effects of friends' selections on user behavior.} (ii) \emph{We propose a social behavioral attention mechanism to adaptively weigh the influence of friends' preferences, and therefore focus on a subset of friends that have similar behaviors when generating recommendations.} In our experiments on benchmark datasets from Epinions and Flixster we evaluate the performance of the proposed NAS model compared to other state-of-the-art methods, as well as we further study the importance of the proposed social behavioral attention mechanism.

The remainder of the paper is organized as follows, Section~\ref{sec:rel} reviews related work, and then Section~\ref{sec:prop} details the proposed NAS model. Finally, in Section~\ref{sec:exp} we examine the performance of the proposed model against baseline models, and Section~\ref{sec:conc} concludes the study.

\section{Related Work}\label{sec:rel}
\subsection{Social Recommendation}
To leverage the recommendation accuracy with friends' selections, Ma et al.~\cite{SOREC} present a social regularization method by sharing a common user-item matrix, factorized by ratings and social relationships. Jamali and Ester~\cite{SOCIALMF} extend~\cite{SOREC} by weighting the user latent factors based on their social relationships. In~\cite{STE}, a social ensemble method is presented to combine matrix factorization with a social-based neighborhood model. Guo et al.~\cite{TRUSTSVD} extend  SVD++~\cite{KDD08} to learn both the user preferences and the social influence of her friends. The aforementioned methods exploit social relationships and use different squared loss functions to minimize the prediction/rating error. On the other hand, several methods focus on the ranking performance when generating the recommendations with social relationships. For example, Grimberghe et al.~\cite{WSDM} combine multi-relational matrix factorization with the Bayesian personalized ranking framework~\cite{BPR} to model users' feedback both on items and on social relationships. Zhao et al.~\cite{CIKM14} propose a social Bayesian personalized ranking model that incorporates social relationships into a pair-wise ranking model, assuming that users tend to assign higher ranks to items that their friends prefer. In~\cite{TW09}, Jamali and Ester combine TrustWalker~\cite{KDD09} with collaborative filtering to generate recommendations with social relationships. In~\cite{RAF16a}, a collaborative ranking strategy is followed, considering how well the relevant items of users and their social friends have been ranked at the top of the list. In~\cite{RAF16b}, authors extend the model presented in~\cite{RAF16a} by combing different collaborative ranking strategies into a joint model to leverage the recommendation accuracy. Chaney et al.~\cite{SPF} infer each user's preferences and the social influence of her friends by introducing a Bayesian model that performs social poisson factorization. 

Recently, a few deep learning strategies have been introduced to generate recommendations with social relationships. For example, Fan et al.~\cite{FanLC18} couple probabilistic matrix factorization with a  feedforward neural network to learn the non-linear features of users' social relationships and their preferences. In a similar spirit, Deng et al.~\cite{DengHXWW17} jointly learn matrix factorization with Denoising Autoencoders, while Nguyen and Lauw~\cite{NguyenL16} study Restricted Boltzmann Machines for representation learning of social behavior. Although these deep learning strategies can compute the non-linearity in friends' preferences, they omit the fact that only a subset of friends have similar behavior. Consequently, these deep learning strategies have limited recommendation accuracy as we will show in Section~\ref{sec:exp}.

\subsection{Neural Attention Models}
More recently, neural attention models have been introduced to produce recommendations. For example, Ebesu et al.~\cite{Eb18} propose collaborative memory networks, where the associative addressing scheme of the memory module acts as a nearest neighborhood model identifying similar users. The attention mechanism learns an adaptive non-linear weighting of the user's neighborhood based on the specific user and item. The output module exploits non-linear interactions between the adaptive neighborhood state jointly with the user and item memories to derive the recommendation. Tay et al.~\cite{Tay18} present a neural attention model to weigh the gravity of users' reviews on items, assuming that not all reviews are created equal, but only a selected few are important. Liu et al.~\cite{Liu18} incorporate attention weights in recurrent neural networks as a priority model to distinguish current interests e.g., clicks from long term preferences. Manotumruksa et al.~\cite{Man18} introduce a contextual attention gate that controls the influence of the ordinary context on the users' contextual preferences and a time- and geo-based gate that controls the influence of the hidden state from the previous check-in based on the transition context. Sun et al.~\cite{Sun18} investigate the problem of how to enhance the temporal recommendation performance, introducing an attentive recurrent network based approach. Attentional factorization machines learn the importance of each feature interaction for content-aware recommendation~\cite{Xiao17}. Nonetheless, all the aforementioned attention-based models try to capture users' sequential behavior or to learn which contextual factors are important when generating recommendations, and do not consider the impact of friends' selections on user behavior.

\begin{table}[t]
\caption{Notation.}
\vspace{-0.2cm}
\begin{center}
\resizebox{\columnwidth}{!}{%
\begin{tabular}{l|l} \hline
\textbf{Symbol} & \textbf{Description} \\ \hline
$n$ & Number of users \\ \hline
$m$ & Number of items \\ \hline
$d$ & Number of latent dimensions \\ \hline
$h$ & Number of hidden layers \\ \hline
$\mathbf{R} \in \mathbb{R}^{n \times m}$ & User-item interaction (rating) matrix \\ \hline
$\mathbf{A} \in \mathbb{R}^{n \times n}$ & Adjacency matrix of users' social relationships \\ \hline
$\mathbf{u}_{u} \in \mathbb{R}^{d \times 1}$ &  Latent vector of user $u$, with $u=1,\ldots,n$ \\ \hline
$\mathbf{v}_{i}  \in \mathbb{R}^{d \times 1}$ & Latent vector of item $i$, with $i=1,\ldots,m$  \\ \hline
$\mathbf{u}_{ufk} \in \mathbb{R}^{d \times 1}$ &  Latent vector of social friend $fk$ of user $u$ \\ \hline
$\mathbf{f}_{up} \in \mathbb{R}^{d \times 1}$ & User's $u$ latent vector based on the social effect of friend $p$, with $p=1,\dots,k$\\ \hline
$\mathbf{z}_{u} \in \mathbb{R}^{d \times 1}$ & Social latent vector of user $u$ \\ \hline
$\mathbf{W}^{e}_{q} \in \mathbb{R}^{d \times d}$ & Weight matrix in the $q$-th hidden layer of model $e$, with $q=1,\ldots, h$\\ \hline
$\mathbf{o}^{e}_{q} \in \mathbb{R}^{d \times 1}$ & Users' hidden representation in the $q$-th hidden layer of model $e$\\ \hline
\end{tabular}}
\end{center}\label{tab:not}
\end{table}

\section{The Proposed NAS Model} \label{sec:prop}
Our notation is presented in Table~\ref{tab:not}. Let $n$ and $m$ be the numbers of users and items, respectively. In matrix $\mathbf{R} \in \mathbb{R}^{n \times m}$, we store user preferences on items, in the form of explicit feedback e.g., ratings or in the form of implicit feedback e.g., number of views, clicks, and so on. By factorizing matrix $\mathbf{R}$, that is, minimizing the approximation error $\sum_{(u,i)} || \mathbf{R}_{ui} - \mathbf{u}_u^\top \mathbf{v}_i ||^2$,  we compute the user latent vectors $\mathbf{u}_u \in \mathbb{R}^{d \times 1}$ and item latent vectors $\mathbf{v}_i \in \mathbb{R}^{d \times 1}$, with $u=1,\ldots,n$,  $i=1,\ldots,m$ and $d$ being the number of latent dimensions. In addition, users' social relationships are stored in the adjacency matrix $\mathbf{A} \in \mathbb{R}^{n \times n}$, if users $a$ and $b$ are friends then $\mathbf{A}_{ab}=1$, and 0 otherwise. Each user $u$ has a set of social friends $\{f_{u1},\ldots,f_{uk}\}$, where $k$ is the number of friends of user $u$. The goal of the proposed NAS model is to weigh the influence of friends' selections on user preferences and generate recommendations, accordingly.

\begin{center}
\begin{figure*} [t]
    \includegraphics[width=\textwidth]{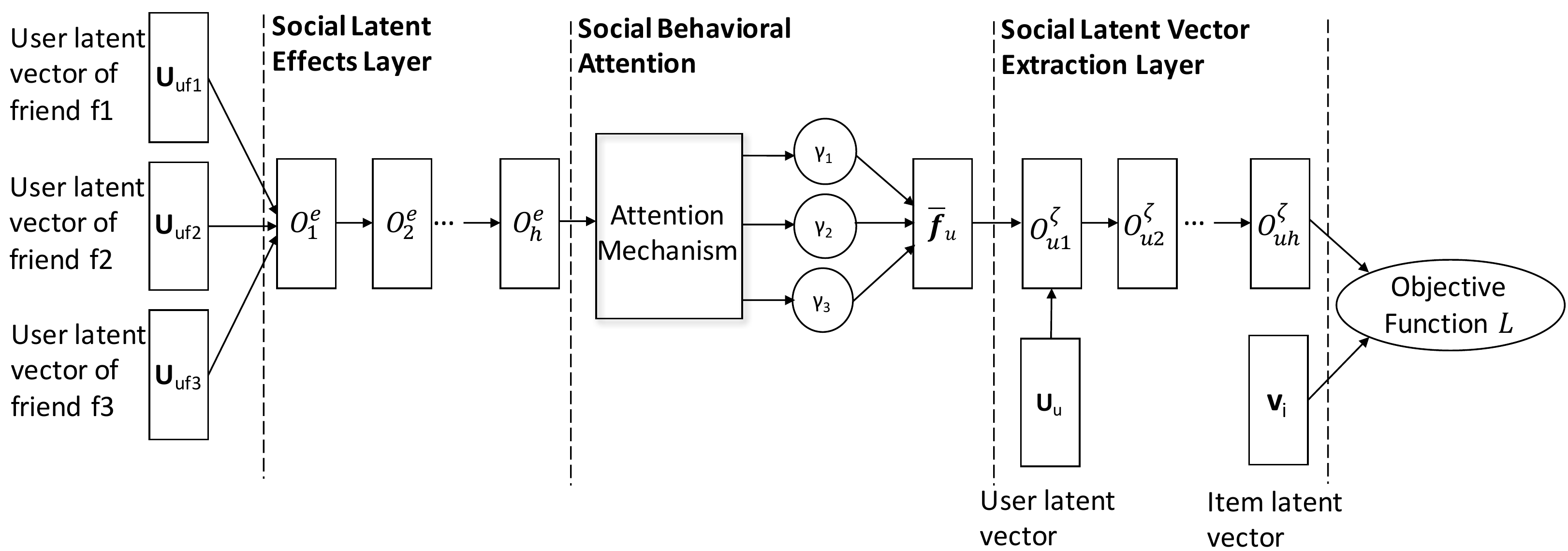}
\caption{An overview of the proposed NAS model. Our architecture consists of the \emph{social latent effects layer}, the \emph{social behavioral attention mechanism} and the \emph{social latent vector extraction layer}. The model parameters are learned based on the objective function $\mathcal{L}$ via backpropagation.} \label{fig:over}    
\end{figure*}
\end{center}

\subsection{NAS Overview}
Figure~\ref{fig:over} illustrates  an overview of the proposed NAS model.  In the example presented in Figure~\ref{fig:over}, we consider $k=3$ social friends of user $u$. By factorizing the user-item matrix $\mathbf{R}$ we compute the user latent vectors $\mathbf{u}_{u}$ and those of their social friends $\mathbf{u}_{ufk}$, as well as the item latent vectors $\mathbf{v}_{i}$ with $u=1,\ldots,n$ and $i=1,\ldots,m$. In the \emph{social latent effects layer}, we compute the social latent effects of the $k$ friends on user's $u $ preferences based on a Multi-Layer Perceptron (MLP) network. Then, in the \emph{social behavioral attention mechanism}, we calculate the attention weights $\gamma_k$ of the latent effects, which correspond to how much users' preferences match with those of their $k$ social friends. In the \emph{social latent vector extraction layer} we capture the non-linear associations between the aggregated social latent effects, that is, vector $\bar{\mathbf{f}}_u \in \mathbb{R}^{d \times 1}$, and the user latent vector $\mathbf{u}_{u}$ via a MLP network. The output is a social user latent vector $\mathbf{o}^{\zeta}_{uh}=\mathbf{z}_{u} \in \mathbb{R}^{d \times 1}$ which is combined with the item latent vector $\mathbf{v}_{i}  \in \mathbb{R}^{d \times 1}$ to learn the objective function $\mathcal{L}$ of our model. 

The remainder of the Section is structured as follows, Section~\ref{sec:eff} presents the social latent effects layer, and Section~\ref{sec:att} details the social behavioral attention mechanism. Finally,  Section~\ref{sec:clv} presents the social latent vector extraction layer, and in Section~\ref{sec:objf} we formulate the objective function of the proposed NAS model, while Section~\ref{sec:det} provides the implementation details.

\subsection{Social Latent Effects Layer} \label{sec:eff}
The goal of the first layer of NAS is to compute the social latent effects. To achieve this, we first design a MLP network to model the $k$ friends' social effects on user preferences. We map the friends' latent vectors $\mathbf{u}_{uf1},\ldots,\mathbf{u}_{ufk}$ and the user latent vector $\mathbf{u}_u$ into a shared embedding layer as follows:

$\forall  p=1,\ldots, k$
\begin{equation}
\mathbf{o}^{e}_{0} = g ( \mathbf{W}^{e}_{u0}\mathbf{u}_u + \mathbf{W}^{e}_{p0} \mathbf{u}_{ufp} + \mathbf{b}^e_0 ) 
\end{equation} 
Matrices $\mathbf{W}^{e}_{u0}$ and $\mathbf{W}^{e}_{p0} \in \mathbb{R}^{d \times d}$  are the weight matrices for the latent vectors of user $u$ and her $p$-th social friend, respectively, and  $\mathbf{b}^e_0 \in \mathbb{R}^{d \times 1} $  is the bias vector. $g(x)=\max(0,x)$ is the Rectifier (ReLU) activation function which is non-saturated. The saturation problem occurs when neurons stop learning and their output is near to either 0 or 1, a problem that might be suffered by the sigmoid and tanh functions~\cite{p30}.  Next, we stack $h$ hidden layers on the top of the embedding layer as shown in Figure~\ref{fig:over}, where the representation $\mathbf{o}^{e}_{q}$ of each hidden layer $q$ is computed as follows:

\begin{equation}
\mathbf{o}^{e}_{q} = g ( \mathbf{W}^{e}_{q} \mathbf{o}^{e}_{q-1} + \mathbf{b}^e_q )
\end{equation}
with $q=1,\ldots, h$

Having computed the representation $\mathbf{o}^{e}_{h}$ of the last hidden layer $h$, to capture the $p$-th friend's social effect on user preferences we calculate user's $u$ latent vector $\mathbf{f}_{up}$ as follows:

\begin{equation} \label{eq:fup}
\mathbf{f}_{up} = \mathbf{W}^{e}_{p} \mathbf{o}^{e}_{h} + \mathbf{b}^e_{p}
\end{equation}
where $\mathbf{W}^{e}_{p} \in \mathbb{R}^{d \times d} $ and $\mathbf{b}^e_{p} \in \mathbb{R}^{d \times 1} $ denote the weight matrix and bias vector of the final user latent vector based on the $p$-th friend's social effect, respectively.
 
\subsection{Social Behavioral Attention} \label{sec:att}
To measure the importance of the $k$ different social effects $\mathbf{f}_{up}$, we propose a \emph{social behavioral attention mechanism}. The attention mechanism learns an adaptive weighting function to focus on a subset of friends that have similar behavior with user $u$. Provided the $k$ social effect vectors $\mathbf{f}_{up}$ based on Equation (\ref{eq:fup}) and the user latent vector $\mathbf{u}_{u}$, we employ a single-layer perceptron to calculate the respective attention score of a social friend $p=1,\dots,k$ as follows:

\begin{equation}
\phi(\mathbf{u}_{u}, \mathbf{f}_{up}) = g(\mathbf{W}^\psi_1\mathbf{u}_{u}+\mathbf{W}^\psi_2 \mathbf{f}_{up} + \mathbf{b}^\psi) 
\end{equation}
where $\mathbf{W}^\psi_1$ and $\mathbf{W}^\psi_2 \in \mathbb{R}^{d \times d}$ are the weight matrices, and $\mathbf{b}^e_{pf} \in \mathbb{R}^{d \times 1} $ is the bias. The superscript $\psi$ refers to the model for the social behavioral attention mechanism.  Then, the final weights are computed by normalizing the respective $k$ social attention scores $\gamma(\mathbf{u}_{u},\mathbf{f}_{up}) $ with the softmax function, which reflects on the importance of the $p$-th friend's social effect on user's $u$ preferences:

$\forall p=1,\ldots, k$
\begin{equation} \label{eq:att}
\gamma(\mathbf{u}_{u},\mathbf{f}_{up}) = \frac{ \exp(\phi(\mathbf{u}_{u}, \mathbf{f}_{up}) }{ \sum_{v=1}^{k} \exp(\phi(\mathbf{u}_{u}, \mathbf{f}_{uv}) }
\end{equation}
Notice that the attention mechanism selectively weighs the user's and friends similarity on preferences based on the social attention scores $\gamma(\mathbf{u}_{u},\mathbf{f}_{up}) $. Having calculated the $k$ social attention scores based on Equation (\ref{eq:att}), the latent vector $\bar{\mathbf{f}}_u$ of aggregated social effect on user's $u$ behavior is computed as follows:

\begin{equation} \label{eq:fu}
\bar{\mathbf{f}}_u = \sum\limits_{p=1}^{k}\gamma(\mathbf{u}_{u},\mathbf{f}_{up})\mathbf{f}_{up}
\end{equation}

\subsection{Social Latent Vector Extraction Layer} \label{sec:clv}
The goal of the \emph{social latent vector extraction layer} is to forecast how the social effects of the $k$ friends' different behaviors influence the user latent representation. Similar to the \emph{social latent effects layer} of Section~\ref{sec:eff}, we first map the aggregated social domain effect $\bar{\mathbf{f}}_u$ of all $k$ friends (Equation (\ref{eq:fu})) and the user's $u$ latent vector $\mathbf{u}_{u}$ into a shared embedding layer, thus constructing a hidden representation $\mathbf{o}^{\zeta}_{u0} \in \mathbb{R}^{d \times 1}$. Then, vector $\mathbf{o}^{\zeta}_{u0}$ is fed to a MLP network of $h$ hidden layers, to capture the non-linear associations of the complex social effects on user's behavior. The output of the MLP network is the \emph{social latent vector} $\mathbf{z}_{u}=\mathbf{o}^{\zeta}_{uh}\in \mathbb{R}^{d \times 1}$, that is, the last hidden representation of the MLP network, as illustrated in Figure~\ref{fig:over}.

\subsection{Objective Function} \label{sec:objf}
The proposed NAS model aims at the ranking performance of the recommendations. Having computed the social latent vector $\mathbf{z}_{u}$ for each user $u$, with $u=1,\dots,n$, we consider the respective item latent vectors $\mathbf{v}_{i}$, with $i=1,\dots,m$. In particular, we define two disjoint sets, a set $\mathcal{I}^+_u$ of observed items that user $u$ has already interacted with, and a set $\mathcal{I}^-_u$ of  unobserved items. For each observed item $i^+\in \mathcal{I}^+_u$, we randomly sample negative/unobserved items $i^-\in \mathcal{I}^-_u$, for each user $u$. According to the Bayesian Personalised Ranking (BPR) criterion~\cite{BPR}, we try to rank the observed items higher than the unobserved ones, having the following loss function:

\begin{equation} \label{eq:loss}
\mathcal{L} = - \sum_{(u, i^+,i^-)} \log \sigma(\hat{\mathbf{R}}_{ui^+} - \hat{\mathbf{R}}_{ui^-})
\end{equation}
where $\sigma(x)=1/\big(1+\exp(-x)\big)$ is the logistic sigmoid function, $\hat{\mathbf{R}}_{ui^+}=\mathbf{z}_{u}^\top  \mathbf{v}_{i+}$ and $\hat{\mathbf{R}}_{ui^-}=\mathbf{z}_{u}^\top \mathbf{v}_{i-}$.

\subsection{Implementation Details} \label{sec:det}
In our implementation we used Tensorflow\footnote{\url{https://www.tensorflow.org}}. We computed the model's parameters, that is, the weight matrices of Sections 3.2-3.4 via backpropagation with stochastic gradient descent, trying to solve the minimization problem of the ranking loss function $\mathcal{L}$ in Equation (\ref{eq:loss}). We employed mini-batch Adam~\cite{p32}, which adapts the learning rate for each parameter by performing smaller updates for frequent and larger updates for infrequent parameters. We set the batch size of mini-batch Adam to 512 with a learning rate of 1e-4. In each backpropagation iteration we performed negative sampling to randomly select a subset $\mathcal{I}^-_u$ of unobserved items as negative instances. 

In addition, to account for the fact that the gradient-based optimization strategy might find a locally - optimal solution of the model's parameter set, we followed a pretraining strategy~\cite{p29}.  We first trained our model with random initializations using only one hidden layer in the MLP networks, employed in our neural architecture. Then, we used the trained parameters as the initialization of our model and varied the number of hidden layers $h$, where we concluded in the optimal number of hidden layers using cross-validation (Section~\ref{sec:param}). The pretraining strategy is very important for our model. To verify this we tested our model without applying the pretraining strategy and we found that there is an average drop of -4.36\% in the model's performance. This observation has been also confirmed by other relevant studies pointing out that the initialization of the model parameters plays a significant role for the model's convergence and performance~\cite{p26}.

\section{Experimental Evaluation} \label{sec:exp}
\subsection{Datasets} \label{sec:data}
In our experiments we used the publicly available datasets, from \emph{Epinions}\footnote{\url{https://alchemy.cs.washington.edu/data/epinions/}} and \emph{Flixster}\footnote{\url{http://www.cs.ubc.ca/~jamalim/datasets/}}. The \emph{Epinions} dataset contains 571,325 ratings on a 5-star scale of 71,002 users on 104,356 items with 508,960 social relationships. The \emph{Flixster} dataset includes 8,196,077 ratings of 147,612 users on 48,794 items with 7,058,819 social relationships. The reason for selecting  both datasets is that they are among the largest publicly available datasets with user preferences and social relationships that have been reported in the relevant literature, reflecting on the real-world sparse setting e.g., the densities of the \emph{Epinions} and \emph{Flixster} datasets are 0.0077 and 0.1138\%, respectively.

\subsection{Evaluation Protocol} We trained the examined models on the 25, 50 and 75\% of the ratings. We used 10\% of the ratings as cross-validation set to tune the models' parameters and evaluate the examined models on the remaining test ratings. To evaluate the top-$N$ recommendation performance of the examined models we used the ranking-based metrics recall and Normalized Discounted Cumulative Gain (NDCG). Recall is defined as the ratio of the relevant items in the top-$N$ ranked list over all the relevant items for each user. The NDCG metric considers the ranking of the relevant items in the top-$N$ list. For each user the Discounted Cumulative Gain is defined as: 
\begin{equation}
 DCG@N = \sum_{l=1}^{N}{\frac{2^{rel_l}-1}{\log_2{l+1}}}
\end{equation}
where $rel_l$ represents the relevance score of item $l$, that is, binary relevance in our case. To remove user rating bias from our results, we considered an item as relevant if a user has rated it above her average ratings and irrelevant otherwise~\cite{Rec17}. $NDCG@N$ is the ratio of $DCG@N$ over the ideal $iDCG@N$ value for each user, that is, the $DCG@N$ value given the ratings in the test set. We set $N$=10 and repeated our experiments five times, and in our results we report average recall and $NDCG$ over the five runs.

\subsection{Compared Methods}
In our experiments we compare the following methods:
\begin{itemize}
\item \emph{\textbf{BPR}}~\cite{BPR}: a baseline ranking model that tries to rank the observed/rated items over the unobserved ones. BPR does not exploit the selections of social friends when generating recommendations. 
\item \emph{\textbf{SoRec}}~\cite{SOREC}: a baseline social-based model that regularizes and weighs friends' latent factors when factorizing the user-item matrix.
\item \emph{\textbf{SBPR}}~\cite{CIKM14}: a social Bayesian ranking model that considers social relationships in the learning process, assuming that users tend to assign higher ranks to items that their friends prefer.
\item \emph{\textbf{DLMF}}~\cite{DengHXWW17}: a deep learning strategy for jointly learning Denoising Encoders with matrix factorization to capture the non-linearity in friends' preferences.
\item \emph{\textbf{DeepSoR}}~\cite{FanLC18}: a deep learning strategy for coupling probabilistic matrix factorization with a feedforward neural network to learn the social interactions of users along with their preferences.
\item \emph{\textbf{NAS*}}: a variant of our model, which ignores the social behavioral attention mechanism of Section 3.3, setting $\gamma(\mathbf{u}_{u},\mathbf{f}_{up})$=1 in Equation~(\ref{eq:att}). This variant serves as baseline to study the impact of the proposed social behavior attention mechanism on our model.
\item \emph{\textbf{NAS}}: the proposed neural attention model for social collaborative filtering.
\end{itemize}
The models' parameters have been determined based on a grid selection strategy and cross-validation, and in our experiments we report the best results. The parameter analysis of our model is presented in Section~\ref{sec:param}.

\subsection{Performance Evaluation}

Figures~2 and 3 present the experimental results in terms of recall and NDCG, respectively when varying the training set size in 25, 50 and 75\%. All the social-based models SoRec, SBPR, DLMF, DeepSoR, NAS* and NAS significantly outperform the baseline  BPR model. This is obtained by exploiting friends' selections when generating recommendations, thus reducing the data sparsity on user preferences. On inspection of the results in Figures~2 and 3, we observe that the competitive deep learning strategies DLMF, DeepSoR and NAS* achieve a higher recommendation accuracy than SoRec and SBPR, as the deep learning strategies can capture well the non-linear associations of friends' preferences in the hidden representations of their neural architectures. Although the three competitive deep learning strategies DLMF, DeepSoR and NAS* have different neural architectures, coupled with matrix factorization, we observe that they achieve comparable performance. In particular, we found that the differences of DLMF, DeepSoR and NAS* in our five runs are not always statistically significant for $p<$0.05 in the paired t-test.

The proposed NAS model achieves a 9.51\% improvement on average in terms of recall when comparing with the second best method, that is, DLMF, DeepSoR or NAS*. Similarly, NAS  outperforms the second best method by an average improvement of 11.12\% in terms of NDCG. Using the paired t-test we found that NAS is superior over all the competitive approaches for $p<$0.05. NAS beats the baselines, as it adaptively selects the weights of friends' preferences. Consequently, the proposed NAS model can self-adjust the subset of friends that express similar behavior, thus producing accurate recommendations. On the other hand, despite that the deep learning strategies DLMR and DeepSoR compute the non-linearity in friends' preferences, a self-tuning weighting strategy on friends' behavior is omitted which explains their limited performance, compared to the proposed NAS model.

To further verify this, we observe that the NAS* variant has limited performance when comparing with the NAS model, having relative drops of -12.57 and -13.85\% on average in terms of recall and NDCG, respectively. This indicates that indeed the social behavioral attention mechanism of the proposed NAS model is a key factor to boost the recommendation accuracy, by adaptively assigning larger weights to the friends that have similar behavior with a user.

\begin{figure*}[t]
\centering
\begin{tabular}{c}  \label{fig:res1}
\hspace{-2cm}\includegraphics[width=15cm]{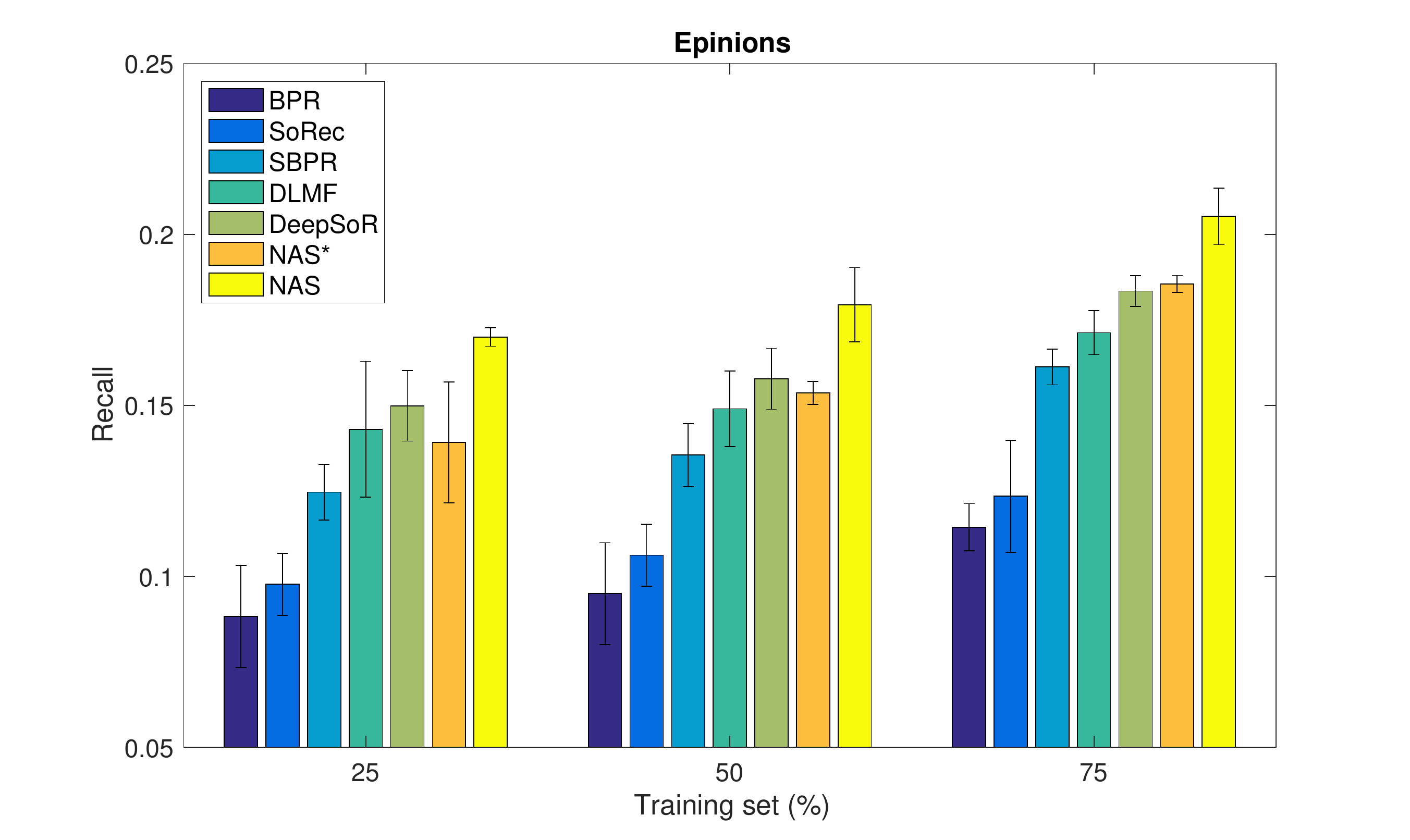} \\ 
\hspace{-2cm}\includegraphics[width=15cm]{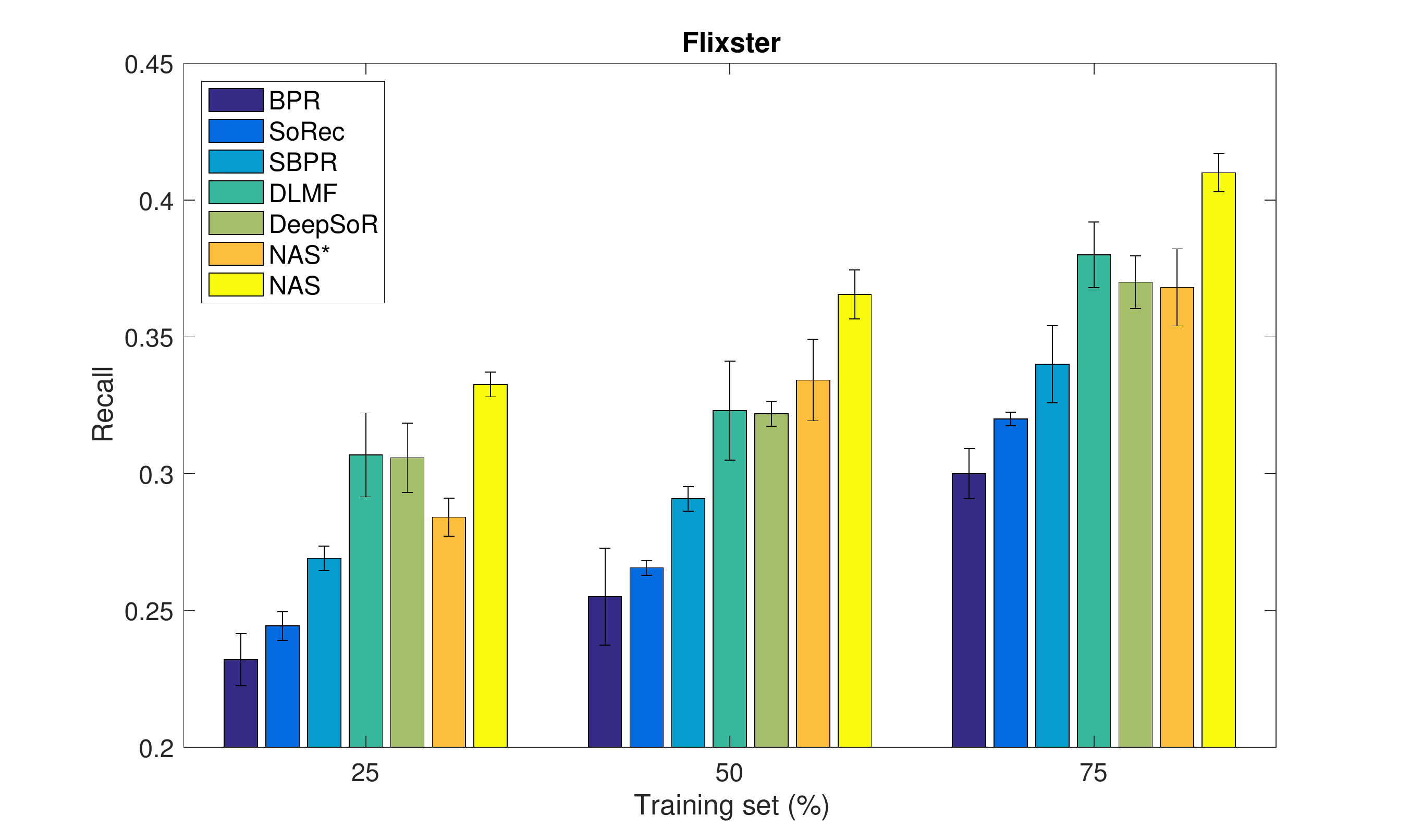}
\end{tabular}
\caption{Performance evaluation in terms of recall for the Epinions and Flixster datasets, when varying the training set size.} 
\end{figure*}

\begin{figure*}[t]
\centering
\begin{tabular}{c}  \label{fig:res2}
\hspace{-2cm}\includegraphics[width=15cm]{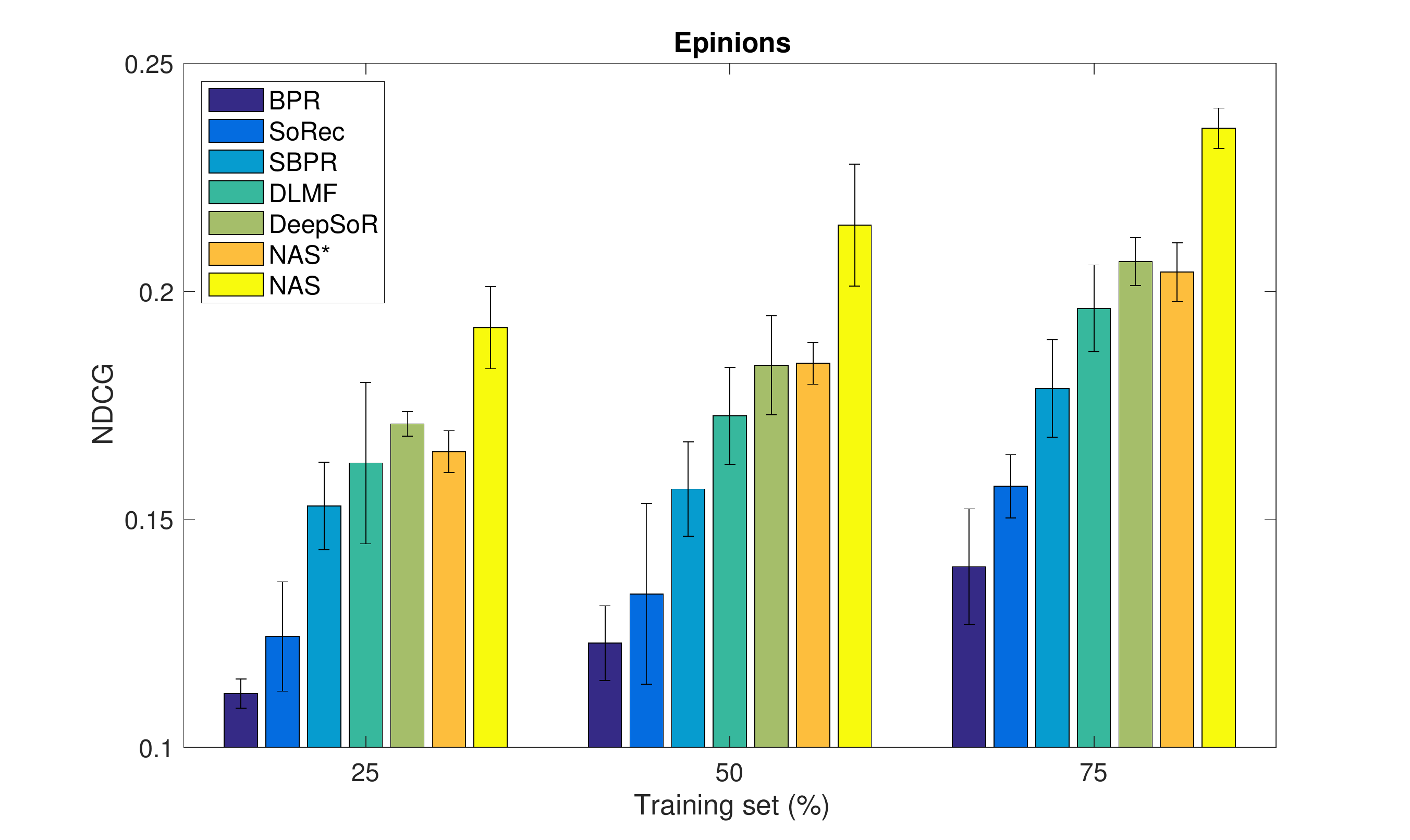} \\ 
\hspace{-2cm}\includegraphics[width=15cm]{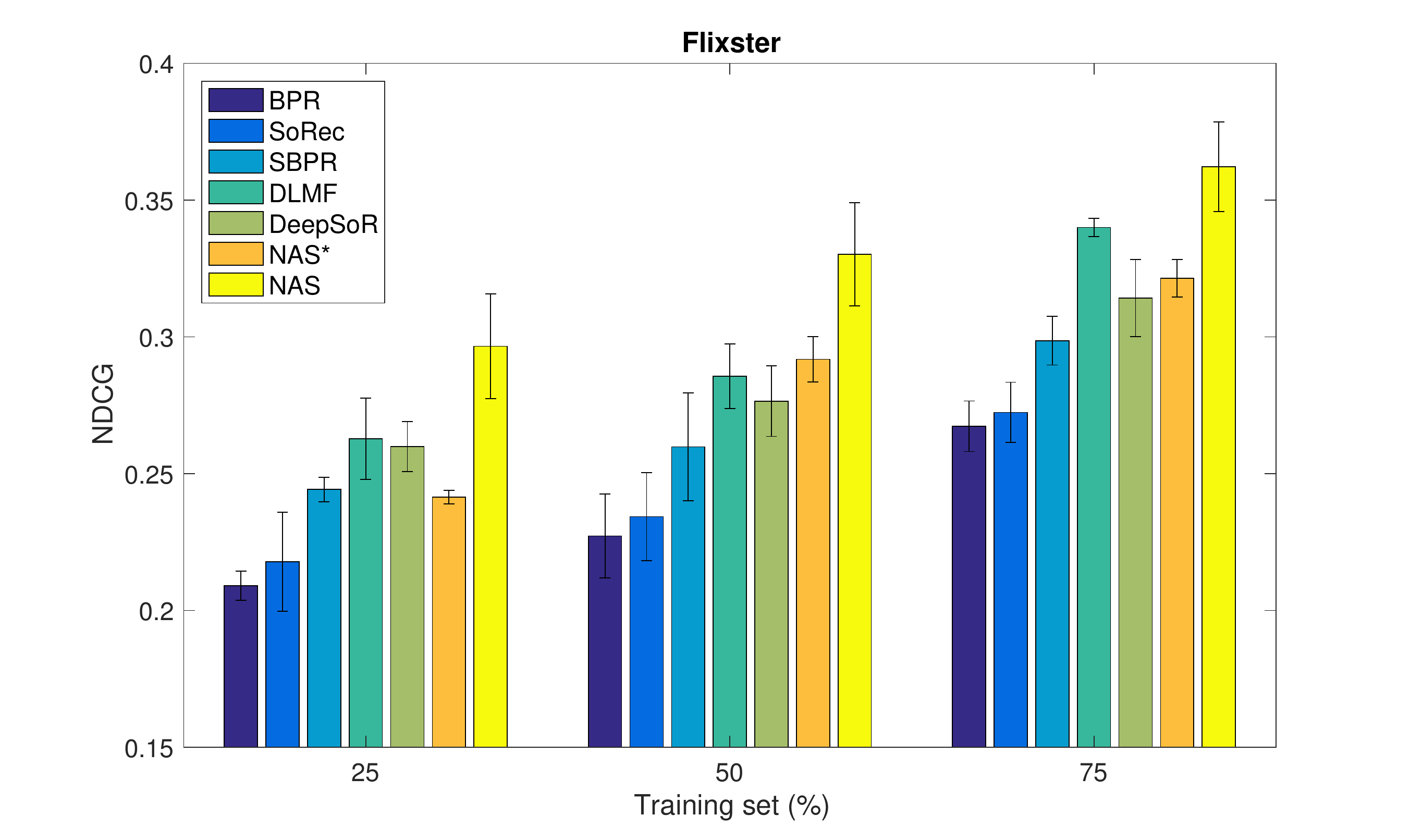}
\end{tabular}
\caption{Performance evaluation in terms of NDCG for the Epinions and Flixster datasets, when varying the training set size.} 
\end{figure*}

\subsection{Parameter Analysis} \label{sec:param}
The three most important parameters in our NAS model are: (i) the number of latent dimensions $d$, (ii) the number of negative samples $|\mathcal{I}^-_u|$ for each observed sample, and (iii) the number of hidden layers $h$. In our implementation we varied the number of latent dimensions $d$ from 10 to 100 by a step of 10 and the number of negative samples $|\mathcal{I}^-_u|$ in 1-10 by a step of 1, for each positive/observed sample. In addition, we tuned the number of hidden layers $h$ from 1 to 5 by a step of 1. Notice that when varying the three parameters we keep them fixed after a point as we observed that they did not significantly increase the recommendation accuracy without paying off in terms of computational cost. 

As illustrated in Figures~4 and 5 more latent dimensions $d$ and hidden layers $h$ are required for the Flixster dataset than the Epinions dataset. This occurs because the Flixster dataset has 14.3 times more ratings and 13.8 times more social relationships than the Epinions dataset (Section~\ref{sec:data}). As a consequence more latent dimensions and hidden layers are required to capture the non-linearity in a larger set of friends' preferences. In our experiments we fix $d$=50 and $h$=3 for the Epinions dataset, and $d$=80 and $h$=4 for the Flixster dataset. 

On the contrary, as Figure~6 shows less negative samples $|\mathcal{I}^-_u|$ are required for the Flixster dataset than the Epinions dataset. This happens because there are more observed ratings in the Flixster dataset, thus less negative samples are needed for each observed sample. In our experiments we set $|\mathcal{I}^-_u|$=9 and 6 for the Epinions and Flixster datasets, respectively.

\begin{figure*}[t]
\centering
\begin{tabular}{cc}  \label{fig:res3}
\hspace{-0.5cm} \includegraphics[width=7cm]{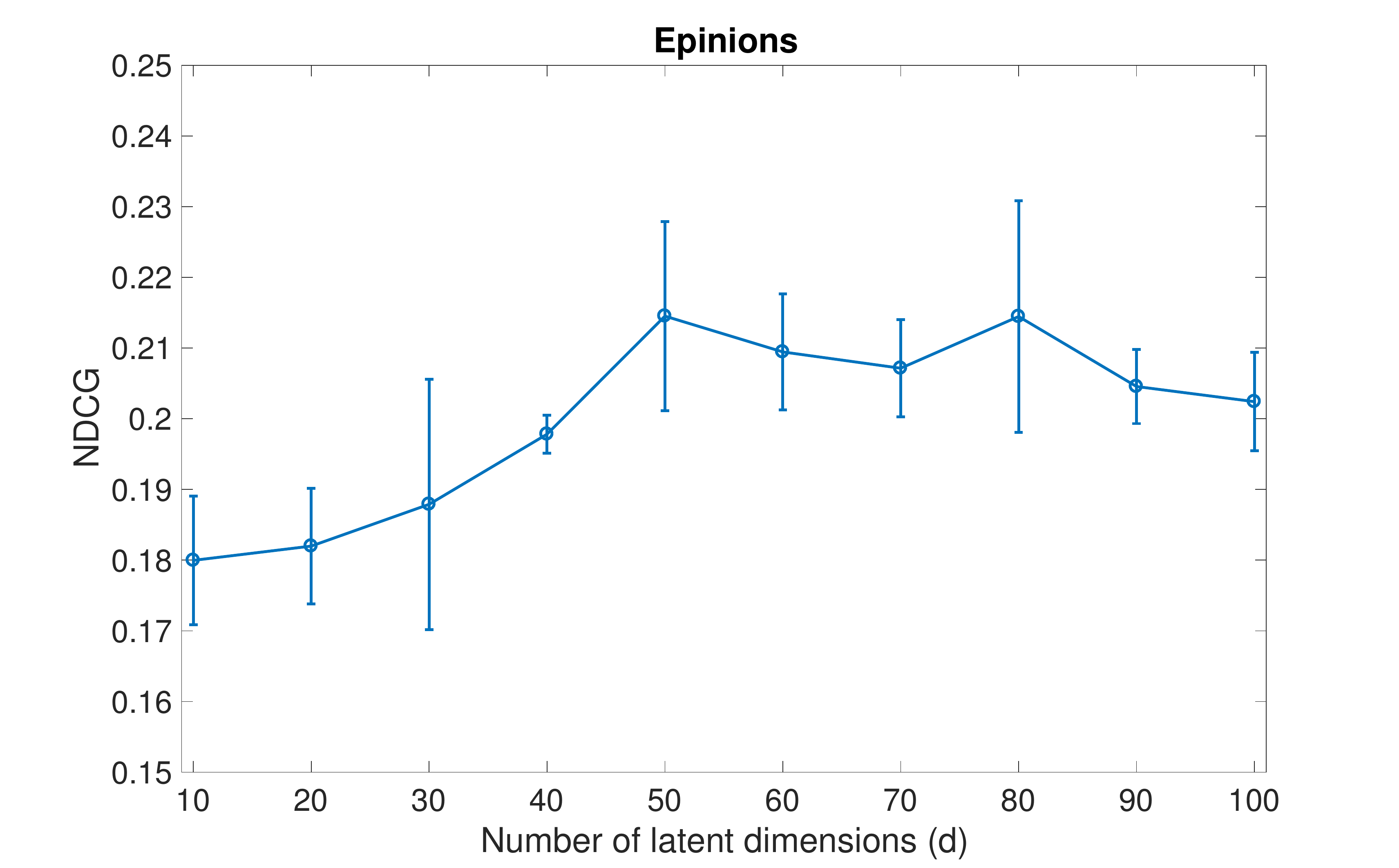} & \hspace{-0.5cm}\includegraphics[width=7cm]{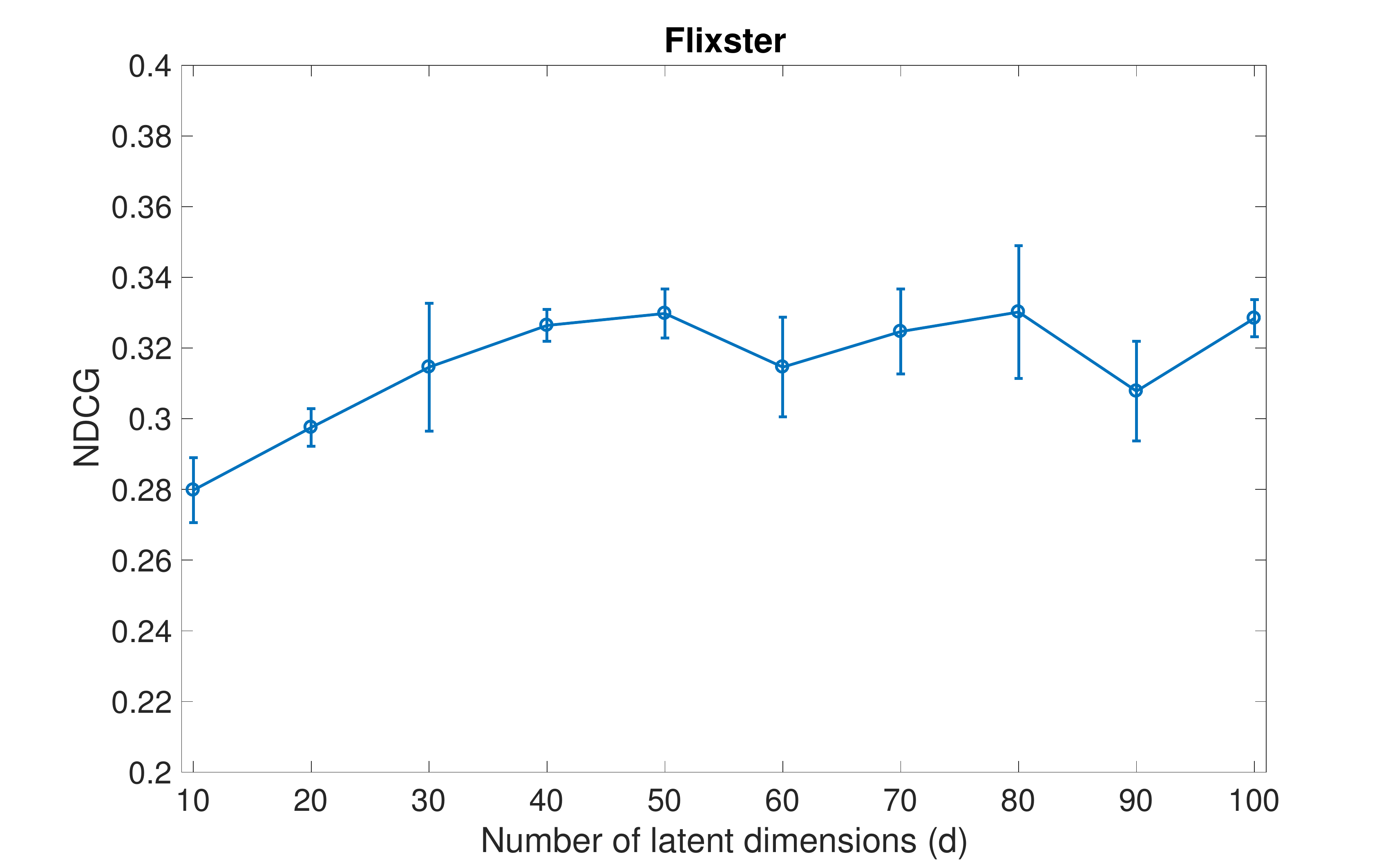}\\ 
\end{tabular}
\caption{Effect on NDCG when varying the number of latent dimensions $d$.} 
\end{figure*}

\begin{figure*}[t]
\centering
\begin{tabular}{cc}  \label{fig:res4}
\hspace{-0.5cm} \includegraphics[width=7cm]{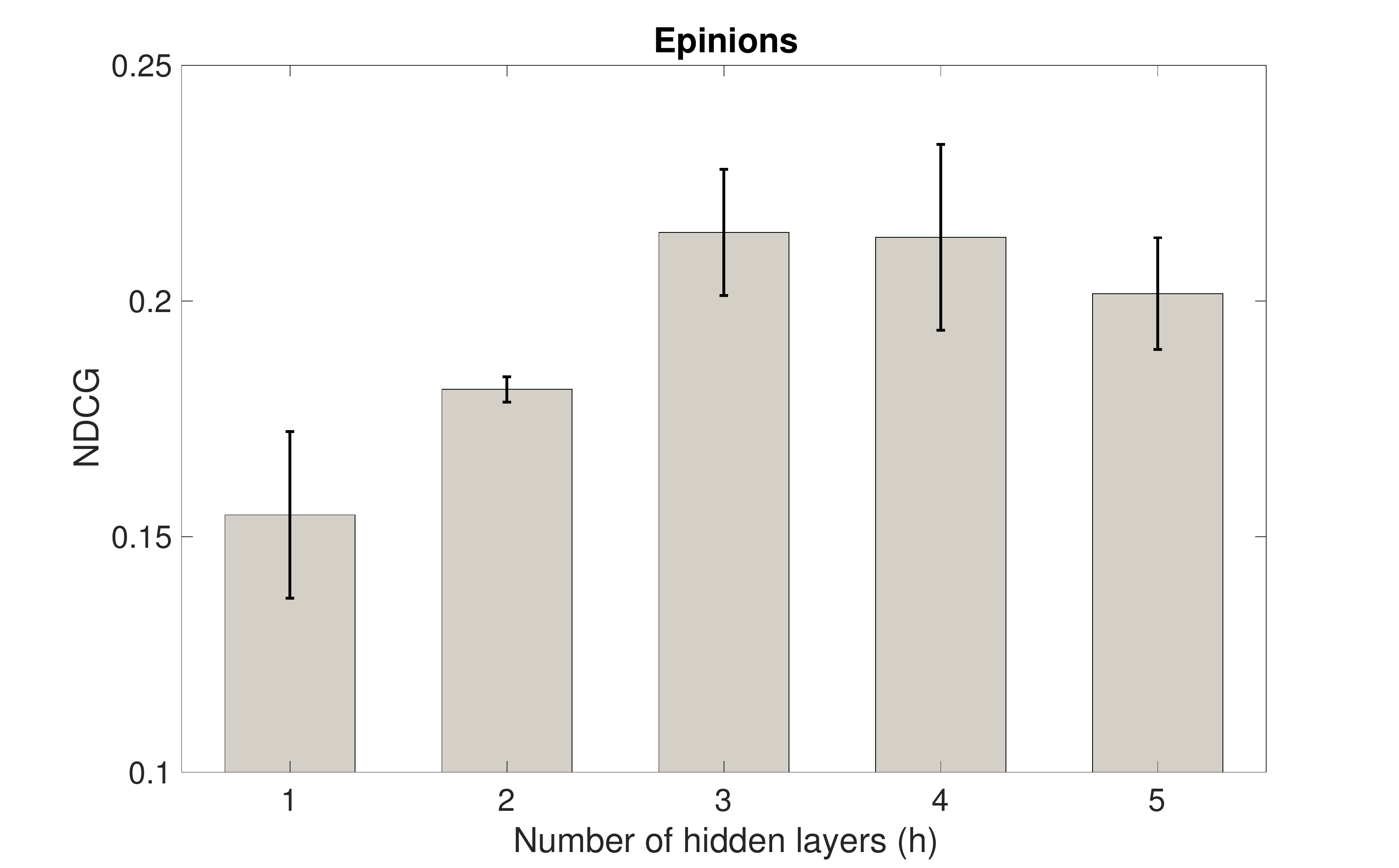} & \hspace{-0.5cm}\includegraphics[width=7cm]{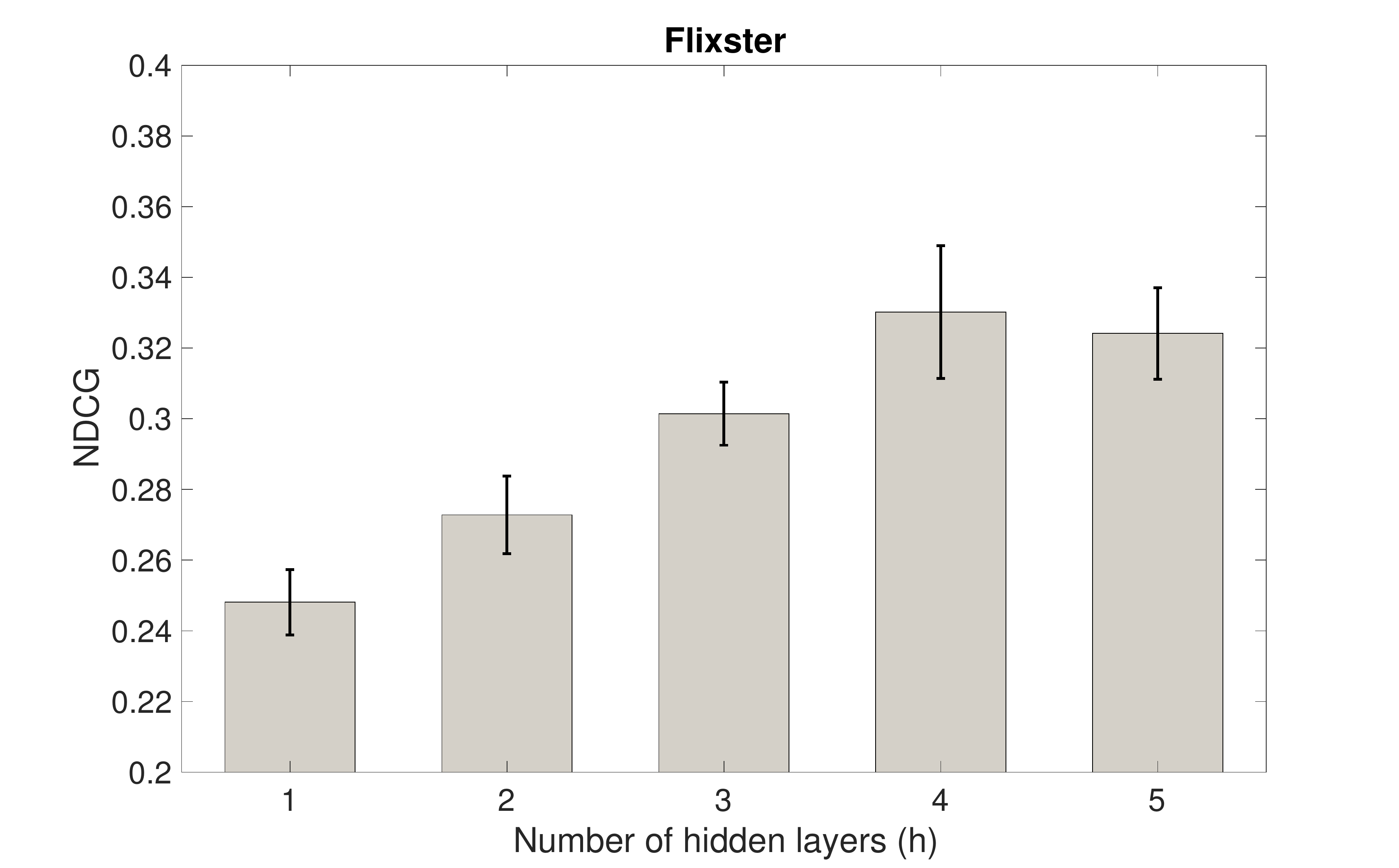}\\ 
\end{tabular}
\caption{Effect on NDCG when varying the number of hidden layers $h$.} 
\end{figure*}

\begin{figure*}[t]
\centering
\begin{tabular}{cc}  \label{fig:res5}
\hspace{-0.5cm} \includegraphics[width=7cm]{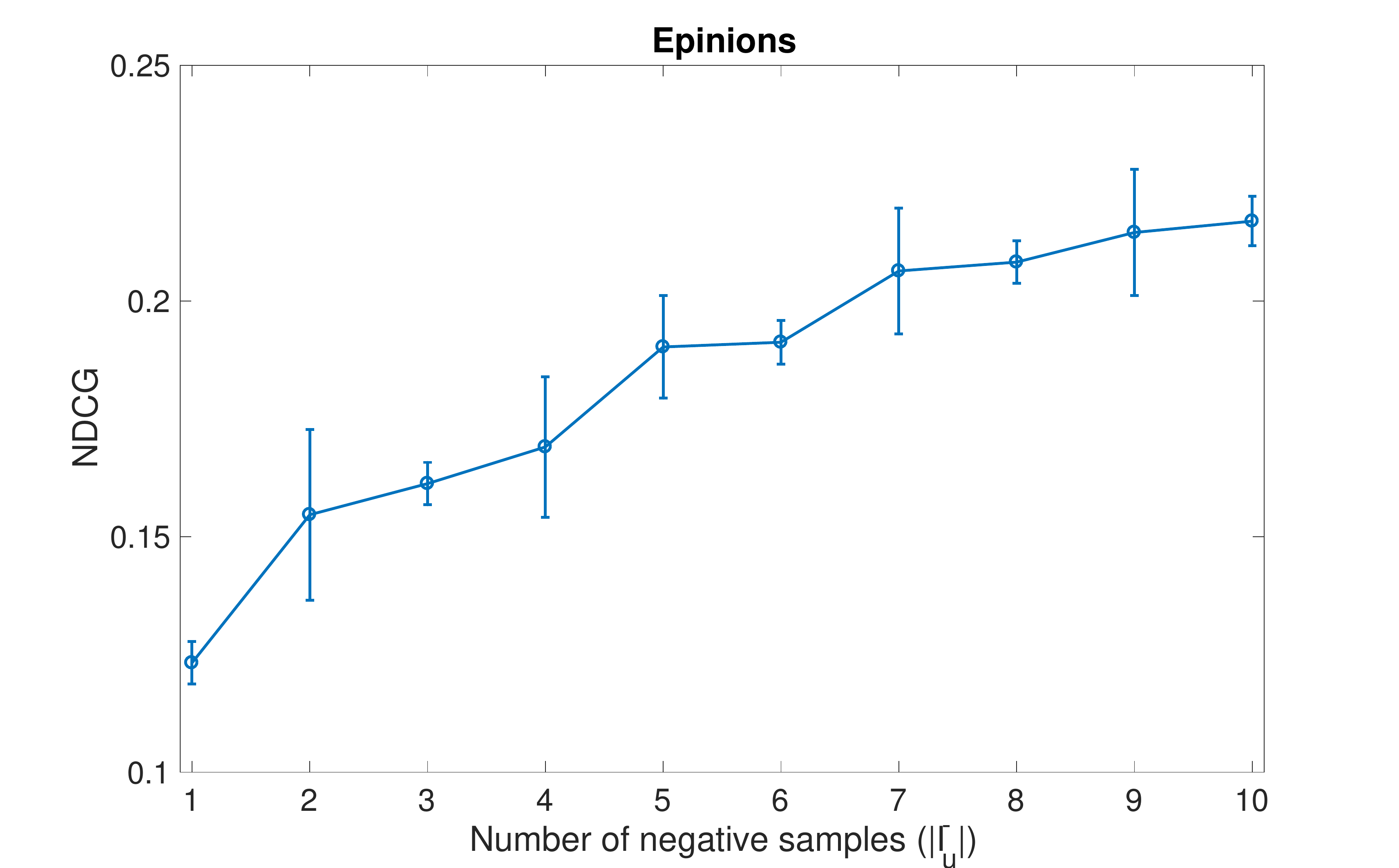} & \hspace{-0.5cm}\includegraphics[width=7cm]{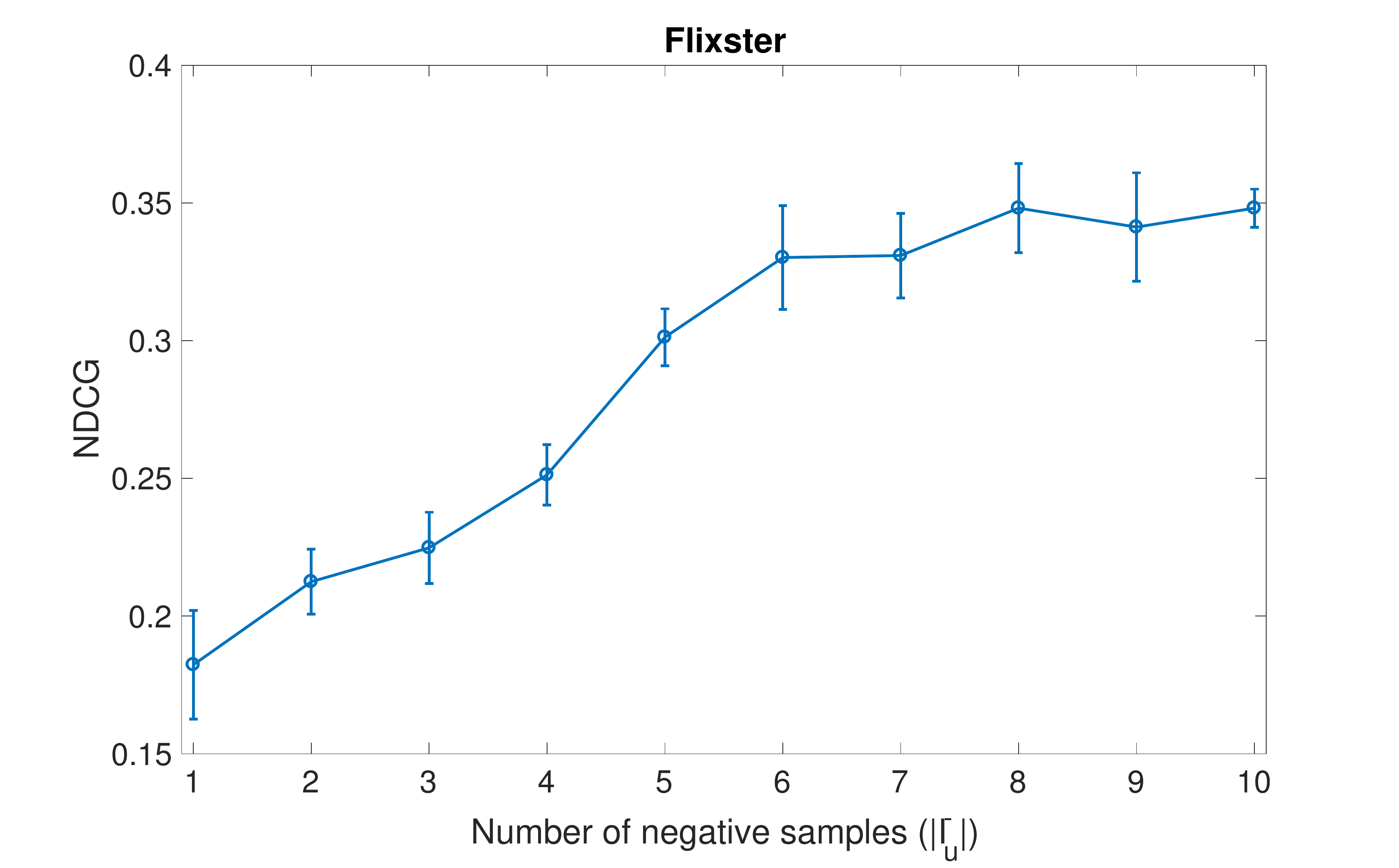}\\ 
\end{tabular}
\caption{Effect on NDCG when varying the number of negative samples $|\mathcal{I}^-_u|$ for each observed sample/rating.} 
\end{figure*}

\section{Conclusions}\label{sec:conc}
In this paper we presented NAS, a neural attention model for social collaborative filtering. The two key factors of our model are (i) to capture the non-linear associations of friends' preferences by computing the social latent effects of friends' preferences on user behavior and (ii) to self-adjust and weigh the influence of friends selections based on a social behavioral attention mechanism. Our experimental evaluation on two publicly available datasets showed the effectiveness of the proposed NAS model, evaluated against other state-of-the-art methods. Compared to the second best method, the proposed NAS model attains an average improvement of 9.51 and 11.12\% in terms of recall and NDCG in all runs. We also evaluated the impact of the proposed social behavioral attention mechanism, using a variant of our model, namely NAS* that omits the proposed attention mechanism. Our experimental results demonstrated that NAS outperformed its variant. Clearly, the social behavioral attention mechanism can significantly boost the recommendation accuracy, by adaptively weighting the impact of friends' selections on user preferences when producing recommendations.

As future work we plan to study how to extend the proposed model to perform cross-domain recommendation~\cite{GAO13,AliannejadiRC17,RafailidisC16,RafailidisC17}. This is a challenging task because we have to transfer and weigh the knowledge of user behaviors across different domains. The key factor of generating cross-domain recommendations is to develop a cross-domain attention mechanism to adaptively perform the weighting of user preferences from the source domains, and consequently generate accurate cross-domain recommendations in a target domain. In addition, an interesting future direction is to exploit the proposed neural attention mechanism for pairwise learning~\cite{SemertzidisRSD15}, social event detection~\cite{RafailidisSLSD13}, information diffusion~\cite{AntarisRN14,RafailidisN15}, exploiting distrust information~\cite{Rafailidis16a} and capturing users' preference dynamics~\cite{RafailidisKM17,RafailidisN15a}. 

\bibliographystyle{unsrt}
\bibliography{attRec}

\end{document}

%% file: preamble.tex
\usepackage{cite}

\usepackage{amssymb,amsfonts,amsthm}
\usepackage[fleqn]{amsmath}

\usepackage{algorithmic}
\usepackage{graphicx}
\usepackage{textcomp}
\usepackage{xcolor}

\usepackage{algorithm}

\usepackage{graphicx}
\usepackage{caption}
\usepackage{subcaption}
\usepackage{url}

\usepackage{balance}

\usepackage{mdframed}

\def\BibTeX{{\rm B\kern-.05em{\sc i\kern-.025em b}\kern-.08em
    T\kern-.1667em\lower.7ex\hbox{E}\kern-.125emX}}

\usepackage{todonotes}

\DeclareCaptionLabelFormat{andtable}{#1~#2  \&  \tablename~\thetable}